\documentclass{article}
\usepackage{spconf,amsmath,graphicx,hyperref}
\usepackage{kotex}
\usepackage{algorithm}
\usepackage{algpseudocode}
\usepackage{tabularx}
\usepackage[table]{xcolor}
\usepackage[dvipsnames]{xcolor}
\usepackage{hhline} 
\usepackage{subcaption}
\usepackage{amsfonts}
\usepackage{cite}
\usepackage{color}
\usepackage{url} 
\usepackage{amssymb}
\usepackage{bm}
\usepackage{stfloats}
\usepackage{enumitem}
\usepackage{booktabs}
\usepackage{bm}
\usepackage{graphicx}
\usepackage{pifont}

\DeclareMathOperator*{\argmin}{arg\,min}

\title{Semantic Pilot Design for Data-Aided Channel Estimation Using a Large Language Model}
\twoauthors
 {Sojeong Park}
	{Pohang University of Science and Technology\\
	Dept. of Electrical Engineering\\
	Pohang, Republic of Korea}
 {Hyun Jong Yang \thanks{Accepted to 2026 IEEE International Conference on Acoustics, Speech and Signal Processing (ICASSP 2026) \\ © 2026 IEEE. Personal use of this material is permitted. Permission from IEEE must be obtained for all other uses, in any current or future media, including reprinting/republishing this material for advertising or promotional purposes, creating new collective works, for resale or redistribution to servers or lists, or reuse of any copyrighted component of this work in other works. \\ \textit{Corresponding author: Hyun Jong Yang}}}
	{Seoul National University\\
	Dept. of Electrical and Computer Engineering\\
    \& Institute of New Media and Communications \\
	Seoul, Republic of Korea}

\begin{document}
%
\maketitle
\begin{abstract}
This paper proposes a semantic pilot design for data-aided channel estimation in text-inclusive data transmission, using a large language model (LLM). In this scenario, channel impairments often appear as typographical errors in the decoded text, which can be corrected using an LLM. The proposed method compares the initially decoded text with the LLM-corrected version to identify reliable decoded symbols. A set of selected symbols, referred to as a semantic pilot, is used as an additional pilot for data-aided channel estimation. To the best of our knowledge, this work is the first to leverage semantic information for reliable symbol selection. Simulation results demonstrate that the proposed scheme outperforms conventional pilot-only estimation, achieving lower normalized mean squared error and phase error of the estimated channel, as well as reduced bit error rate.
\end{abstract}
\vspace{-2mm}
\section{Introduction}

Accurate channel state information (CSI) is essential for reliable wireless communication \cite{perfect_csi}. The most common approach to obtain CSI is pilot-based channel estimation, where a pilot sequence is transmitted along with the data \cite{robust_ce, ce1, ce2}. However, the estimation accuracy highly relies on the length of the pilot sequence. Although extending the pilot length enhances channel estimation performance, it also introduces additional overhead.
To address this limitation, data-aided channel estimation has been proposed \cite{da1, da2, da3, da4, da5}. In this approach, a pilot-based initial estimate is used to decode data symbols, which are then exploited to refine the channel estimation. This method improves estimation accuracy without incurring additional pilot overhead. Its performance depends on the reliability of the decoded symbols, as erroneous symbols degrade the estimation accuracy. Therefore, selecting only reliable symbols is essential, and many studies have been conducted on reliable symbol selection techniques \cite{rs1, rs2, rs3}. Although these studies provide effective methods, their scope is often confined to statistical information from the physical layer. This limitation opens up opportunities to explore the contextual and semantic aspects of the data.

In this paper, we propose a channel estimation framework that integrates semantic information via a large language model (LLM). While LLMs have been applied to various communication tasks \cite{llm1, llm2, llm3, llm4}, their potential in channel estimation remains largely unexplored.
In text-inclusive transmission scenarios, typographical errors caused by channel impairments can be corrected by LLMs. Leveraging this capability, the proposed method establishes a semantic pilot by identifying highly reliable symbols via LLM-based correction. This semantic pilot is then applied to refine the channel estimate, effectively enhancing performance without additional pilot overhead. As far as we are aware, no prior work has explored the use of semantic information to support symbol decisions in channel estimation.

\vspace{-2mm}
\section{System Model}
\begin{figure*}[t]
    \centering
    \includegraphics[width=17.5cm]{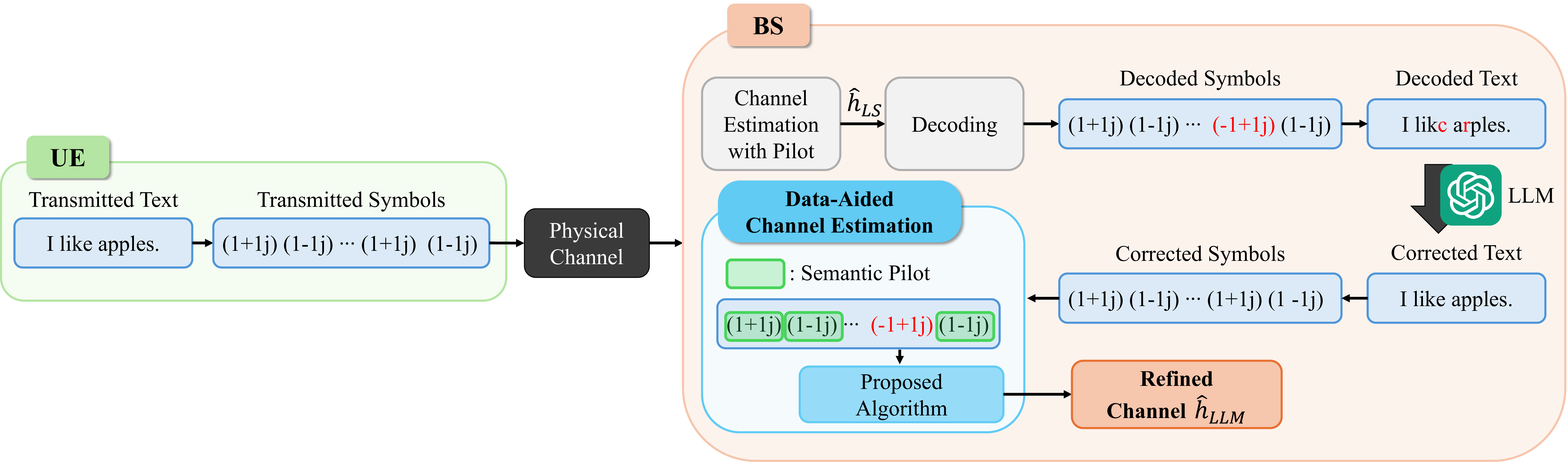}
    \vspace{0pt}
    \caption{System model of the proposed semantic pilot data-aided channel estimation with an LLM.}
    \label{fig:system model}
    \vspace{-3mm}
\end{figure*}
This work considers an uplink single-input single-output (SISO) system, where both the user equipment (UE) and the base station (BS) are equipped with a single antenna. As illustrated in Fig. \ref{fig:system model}, the UE sends text data to the BS through a physical channel. This text is assumed to be transmitted along with other data within the same payload. At the BS, an initial channel estimate is obtained using a pilot sequence and subsequently used to decode the received signal. Imperfect channel estimation and noise typically result in typographical errors in the initially decoded text, which can be corrected using an LLM at the BS. By applying LLM-aided text correction, the proposed method identifies reliable symbols and refines the channel through data-aided estimation. This refined channel is applied to decode the entire payload.

In particular, the UE transmits a text sequence denoted by $\mathbf{t}= [t^{(1)}, t^{(2)}, \dots, t^{(L)}]$, where $t^{(i)}$ represents the $i$-th character in the sequence and $L$ is the length of the transmitted text. This sequence is mapped to a modulated symbol vector $\mathbf{x}_t$, which is transmitted along with a pilot sequence $\mathbf{x}_p$. The overall transmitted signal is given by $\mathbf{x}=[\mathbf{x}_p, \mathbf{x}_t]$. The corresponding received signal at the BS is represented as $\mathbf{y}=[\mathbf{y}_p, \mathbf{y}_t]$, where $ \mathbf{y}_p = h\mathbf{x}_p + \mathbf{n}_p, \mathbf{y}_t = h\mathbf{x}_t + \mathbf{n}_t$.
Therefore, the overall received signal can be compactly written as:
\begin{equation}
    \mathbf{y} = h\mathbf{x} + \mathbf{n},
\end{equation}
where $h$ denotes the channel coefficient between UE and BS, and $\mathbf{n}_p,\mathbf{n}_t,$ and $\mathbf{n}$ represent additive white Gaussian noise, each following $\mathcal{CN}(0, \sigma^2\mathbf{I})$. 

At the BS, the initial channel estimation is performed using the pilot sequence $\mathbf{x}_p$. In this system, we adopt the least squares (LS) method to estimate the channel coefficient, denoted by $\hat{h}_{LS}$. The LS estimate is given by:
\begin{equation} \label{eq:LS}
    \hat{h}_{LS} = \frac{\mathbf{x}_p^H\mathbf{y}_p}{\lVert\mathbf{x}_p\rVert^2},
\end{equation}
where $(\cdot)^H$ denotes the conjugate transpose operator. The estimated channel $\hat{h}_{LS}$ is used to decode the received text data symbols. Specifically, channel equalization is performed using a zero-forcing approach. The equalized symbol vector is obtained as:
\begin{equation}
    \tilde{\mathbf{x}}_t = \frac{\mathbf{y}_t}{\hat{h}_{LS}}.
\end{equation}
Symbol decisions are applied to obtain decoded symbols \( \hat{\mathbf{x}}_t \), which are subsequently demodulated and converted into a text sequence $\hat{\mathbf{t}} = [\hat{t}^{(1)}, \hat{t}^{(2)}, \dots, \hat{t}^{(L)}]$. To address decoding errors caused by channel imperfections, the LLM produces a corrected text sequence $\hat{\mathbf{t}}_{LLM} = [\hat{t}_{LLM}^{(1)}, \hat{t}_{LLM}^{(2)}, \dots, \hat{t}_{LLM}^{(L)}]$, which is then re-encoded into symbols denoted as $\hat{\mathbf{x}}_{LLM}$.

To improve the accuracy of channel estimation, we adopt a data-aided channel estimation approach. In particular, we introduce a symbol selection function $\mathcal{S}(\cdot, \cdot)$, which identifies reliable symbols based on the initially decoded text sequence $\hat{\mathbf{t}}$ and the LLM-corrected text sequence $\hat{\mathbf{t}}_{LLM}$. The selected symbols, assumed to be error-free, are used as additional pilot symbols to refine the channel estimate. The output of this selection is denoted by:
\begin{equation}
\mathbf{x}_s = \mathcal{S}(\hat{\mathbf{t}}, \hat{\mathbf{t}}_{LLM}),    
\end{equation}
and referred to as the semantic pilot. Further details on the selection process are provided in Section~\ref{sec:semantic_pilot}. Using the pilot sequence and the semantic pilot, we refine the channel estimation through a data-aided channel estimation function $\mathcal{R}(\cdot, \cdot)$, which yields a refined channel $\hat{h}_{LLM}$ as:
\begin{equation}
    \hat{h}_{LLM} = \mathcal{R}(\mathbf{x}_p, \mathbf{x}_s).
\end{equation}
The details of the data-aided channel estimation are provided in Section~\ref{sec:data-aided}.

\vspace{-2mm}
\section{Semantic Pilot Design} \label{sec:semantic_pilot}

\begin{table*}[t]
\caption{Example of semantic pilot selection by comparing the initially decoded and LLM-corrected text. Green checkmarks (\textcolor{Green}{\ding{51}}) indicate matching characters used as the semantic pilot. Errors are highlighted in \textcolor{red}{red}.}
\centering
\renewcommand{\arraystretch}{1.2}
\setlength{\tabcolsep}{2pt}
\begin{tabular}{|c|*{32}{c|}}
\hline
\textbf{Transmitted ($\textbf{t}$)} &W &e &\;\; &n &e &e &d &\;\;  &t &o &\;\; &s &p &e &l &l &\;\; &o &u &t &\;\; &t &h &e &\;\; &f &a &c &t &s & \dots \\
\hline
\textbf{Initially Decoded ($\hat{\textbf{t}}$)} &\textcolor{red}{U} &\textcolor{red}{i} &\;\; &\textcolor{red}{X} &e &e &\textcolor{red}{f} &\;\;  &t &\textcolor{red}{E} &\;\; &\textcolor{red}{k} &p &e &\textcolor{red}{V} &l &\;\; &\textcolor{red}{w} &u &t &\;\; &\textcolor{red}{l} &h &\textcolor{red}{W} &\textcolor{red}{'} &\textcolor{red}{j} &a &\textcolor{red}{U} &t &\textcolor{red}{M} & \dots \\
\hline
\textbf{LLM-Corrected ($\hat{\textbf{t}}_{LLM}$)} &W &e &\;\; &n &e &e &d &\;\;  &t &o &\;\; &s &p &e &l &l &\;\; &o &u &t &\;\; &\textcolor{red}{X} &\textcolor{red}{X} &\textcolor{red}{X} &\textcolor{red}{X} &\textcolor{red}{X} &\textcolor{red}{X} &\textcolor{red}{X} &\textcolor{red}{X} &\textcolor{red}{X} & \dots \\
\hline
\textbf{Semantic Pilot ($\textbf{x}_s$)} &- &- &\textcolor{Green}{\ding{51}} &- &\textcolor{Green}{\ding{51}} &\textcolor{Green}{\ding{51}} &- &\textcolor{Green}{\ding{51}}  &\textcolor{Green}{\ding{51}} &- &\textcolor{Green}{\ding{51}} &- &\textcolor{Green}{\ding{51}} &\textcolor{Green}{\ding{51}} &- &\textcolor{Green}{\ding{51}} &\textcolor{Green}{\ding{51}} &- &\textcolor{Green}{\ding{51}} &\textcolor{Green}{\ding{51}} &\textcolor{Green}{\ding{51}} &- &- &- &- &- &- &- &- &- & \dots \\
\hline
\end{tabular}
\label{tab:semantic_pilot_selection}
\vspace{-3mm}
\end{table*}

In data-aided channel estimation, the decoded data symbols are utilized as an additional pilot to enhance channel estimation accuracy. However, inaccurately decoded symbols degrade performance, it is essential to select only reliable symbols as the additional pilot. We define a semantic pilot as a set of reliable decoded symbols derived from the alignment between the initially decoded and LLM-corrected texts. This section describes the LLM prompt for text correction and the semantic pilot selection.

\vspace{-2mm}
\subsection{Text-Correction LLM Prompting}
To correct typographical errors in the decoded text, we prompt an LLM with strict constraints to ensure symbol-level alignment with the transmitted sequence. The corrected text must be the same length as the input, allowing only character substitutions. Severely corrupted segments are replaced with a sequence of `X' characters of equal length. An example of this error correction is shown in Table~\ref{tab:semantic_pilot_selection}.
\vspace{-2mm}
\subsection{Semantic Pilot Selection}
A semantic pilot is determined by comparing the initially decoded text with its LLM-corrected version. If a character remains unchanged after correction, it is regarded as correctly decoded, and its corresponding modulated symbols are considered reliable. The set of reliable symbols is selected as a semantic pilot. An illustrative example is provided in Table~\ref{tab:semantic_pilot_selection}.
Let $\hat{\mathbf{t}}$ denote the initially decoded text and $\hat{\mathbf{t}}_{LLM}$ be its LLM-corrected version. For each character $\hat{t}^{(i)}$, let $\mathcal{M}(\hat{t}^{(i)})$ represent the modulated symbols associated with that character. We define a symbol selection function $\mathcal{S}(\cdot, \cdot)$ that identifies the semantic pilot as:
\begin{equation}
    \begin{split}
    \mathbf{x}_s &= \mathcal{S}(\hat{\mathbf{t}}, \hat{\mathbf{t}}_{LLM}) \\
    &= \{x_s|\; \exists i \text{\;s.t.\;} x_s \in \mathcal{M}(\hat{t}^{(i)}), \; \hat{t}^{(i)} = \hat{t}_{LLM}^{(i)}\}.   
    \end{split}
\end{equation}
The semantic pilot $\mathbf{x}_s$ is subsequently used in data-aided channel estimation.

\vspace{-2mm}
\section{Data-Aided Channel Estimation} \label{sec:data-aided}

To improve estimation accuracy, we propose a data-aided channel estimation method that exploits both the pilot sequence $\mathbf{x}_p = [x_p^{(1)}, x_p^{(2)}, \dots, x_p^{(M)}]$ and the semantic pilot $\mathbf{x}_s = [x_s^{(1)}, x_s^{(2)}, \dots, x_s^{(N)}]$. In the proposed algorithm, we refine the channel estimate in two steps: phase estimation and magnitude scaling.

\vspace{-2mm}
\subsection{Phase Refinement}
The phase of the channel is first refined using LS estimation with the pilot sequence and the semantic pilot. This problem can be formulated as:
\begin{equation}
    \hat{h}_r 
    = \argmin_{h \in \mathbb{C}}
    \Bigg(
        \sum_{i=1}^{M} \big|y_p^{(i)} - h x_p^{(i)} \big|^2
        + \sum_{j=1}^{N} \big|y_s^{(j)} - h x_s^{(j)} \big|^2
    \Bigg),
\end{equation}
where $h \in \mathbb{C}$ denotes the channel coefficient to be estimated, and $y_p^{(i)}$ and $y_s^{(j)}$ are the received symbols corresponding to the pilot symbol and semantic pilot symbol $x_p^{(i)}$ and $x_s^{(j)}$, respectively. The solution is given by:
\begin{equation} \label{eq:phase}
    \hat{h}_r 
    =
    \frac{
        \sum_{i=1}^{M} x_p^{(i)*} y_p^{(i)}
        + \sum_{j=1}^{N} x_s^{(j)*} y_s^{(j)}
    }
    {
        \sum_{i=1}^{M} \big| x_p^{(i)} \big|^2
        + \sum_{j=1}^{N} \big| x_s^{(j)} \big|^2
    },
\end{equation}
where $(\cdot)^*$ denotes the complex conjugate operator.

\vspace{-2mm}
\subsection{Magnitude Scaling}
In constant-envelope modulations, the received magnitude depends mainly on the channel gain and noise. While phase estimation demands the reliable semantic pilot, magnitude estimation leverages the law of large numbers. Therefore, we utilize the entire decoded payload to average out zero-mean noise and robustly correct the magnitude bias while preserving the refined phase.
This is achieved by introducing a scaling factor $\gamma \in \mathbb{R}$, obtained by solving the following problem:
\begin{equation} \label{eq:scaling}
\gamma
= \argmin_{\gamma \in \mathbb{R}} \Bigg(
     \sum_{i=1}^{M} \big| y_p^{(i)} - \gamma \hat{h}_r x_p^{(i)} \big|^2 
     + \sum_{j=1}^{K} \big| y_t^{(j)} - \gamma \hat{h}_r \hat{x}_t^{(j)} \big|^2 \Bigg),
\end{equation}
where $\mathbf{y}_t = [y_t^{(1)}, y_t^{(2)}, \dots, y_t^{(K)}]$ denotes the received text‑data symbols, $\hat{\mathbf{x}}_t = [\hat{x}_t^{(1)}, \hat{x}_t^{(2)}, \dots, \hat{x}_t^{(K)}]$ denotes the corresponding decoded symbols, and $K$ is the length of the text-data symbols.
The solution for $\gamma$ is:
\begin{equation} 
    \gamma
    =
    \frac{
        \Re \Big\{
            \sum_{i=1}^{M} 
                (\hat{h}_r x_p^{(i)})^* y_p^{(i)}
            + \sum_{j=1}^{K} 
                (\hat{h}_r \hat{x}_t^{(j)})^* y_t^{(j)}
        \Big\}
    }
    {
        \sum_{i=1}^{M} 
            \big| \hat{h}_r x_p^{(i)} \big|^2
        + \sum_{j=1}^{K} 
            \big| \hat{h}_r \hat{x}_t^{(j)} \big|^2
    },
\end{equation}
where $\Re\{\cdot\}$ denotes the real part operator. 
Finally, the magnitude-scaled channel estimate is given by $\hat{h}_{LLM} = \gamma \hat{h}_r$.

\vspace{-2mm}
\section{Simulation Results}
\subsection{Simulation Settings}
To evaluate the proposed model, we use the Europarl dataset \cite{koehn-2005-europarl} as the text data. The text is encoded using 6‑bit fixed‑length source coding, and modulated with quadrature phase shift keying (QPSK). Zadoff-Chu sequence of length 16 is employed as the pilot. For text correction, we utilize OpenAI's o4-mini model as an LLM, which is tailored for the task using prompt engineering. All experiments are conducted in a SISO setting over a Rician fading channel.
The following channel estimation schemes are compared:

\begin{itemize}[itemsep=-1.5mm, topsep=0pt]
    \item \textbf{Pilot}: Conventional pilot-only estimation using (\ref{eq:LS}).
    \item \textbf{Decoded}: Conventional data-aided estimation using all initially decoded symbols. The magnitude scaling step is omitted as its scaling factor is mathematically 1.
    \item \textbf{LLM-Corrected}: Data-aided estimation using all LLM-corrected symbols with magnitude scaling.
    \item \textbf{Proposed (w/o Scaling)}: Data-aided channel estimation using the semantic pilot without magnitude scaling, as defined in (\ref{eq:phase}).
    \item \textbf{Proposed}: Data-aided channel estimation using the semantic pilot with magnitude scaling. 
\end{itemize}
\vspace{-2mm}
\subsection{Evaluation Metrics}

First, to analyze the effectiveness of the semantic pilot selection, we define two metrics.
\begin{itemize}[itemsep=-1.5mm, topsep=0pt]
    \item \textbf{Reliability} quantifies the accuracy of the semantic pilot selection, defined as the ratio of error-free semantic pilot symbols to the total number of selected symbols.

    \item \textbf{Detection Rate} reflects how thoroughly reliable symbols are identified, defined as the ratio of error-free semantic pilot symbols to the total number of error-free decoded symbols before LLM correction.

    \item \textbf{Selection Ratio} measures the proportion of the data payload utilized as an additional pilot, defined as the ratio of the number of selected semantic pilot symbols to the total number of data symbols.
\end{itemize}
Additionally, to evaluate the performance of the proposed scheme, we consider three metrics: normalized mean squared error (NMSE), phase error, and bit error rate (BER). Let $h$ denote the true channel and $\hat{h}$ the estimated channel.

\begin{itemize}[itemsep=-1.5mm, topsep=0pt]
    \item \textbf{NMSE} measures the accuracy of the estimated channel and is defined as $\frac{| h - \hat{h}|^2}{| h|^2}$. 

    \item \textbf{Phase error} quantifies the angular difference between the estimated and true channels, given by $| \angle h - \angle \hat{h} |$, where $\angle(\cdot)$ denotes the phase of a complex number.

    \item \textbf{BER} is the ratio of erroneously detected bits to the total number of transmitted bits.
\end{itemize}

\begin{figure}[t!]
    \centering
    \begin{subfigure}[t]{0.48\linewidth}
        \centering
        \includegraphics[width=\linewidth]{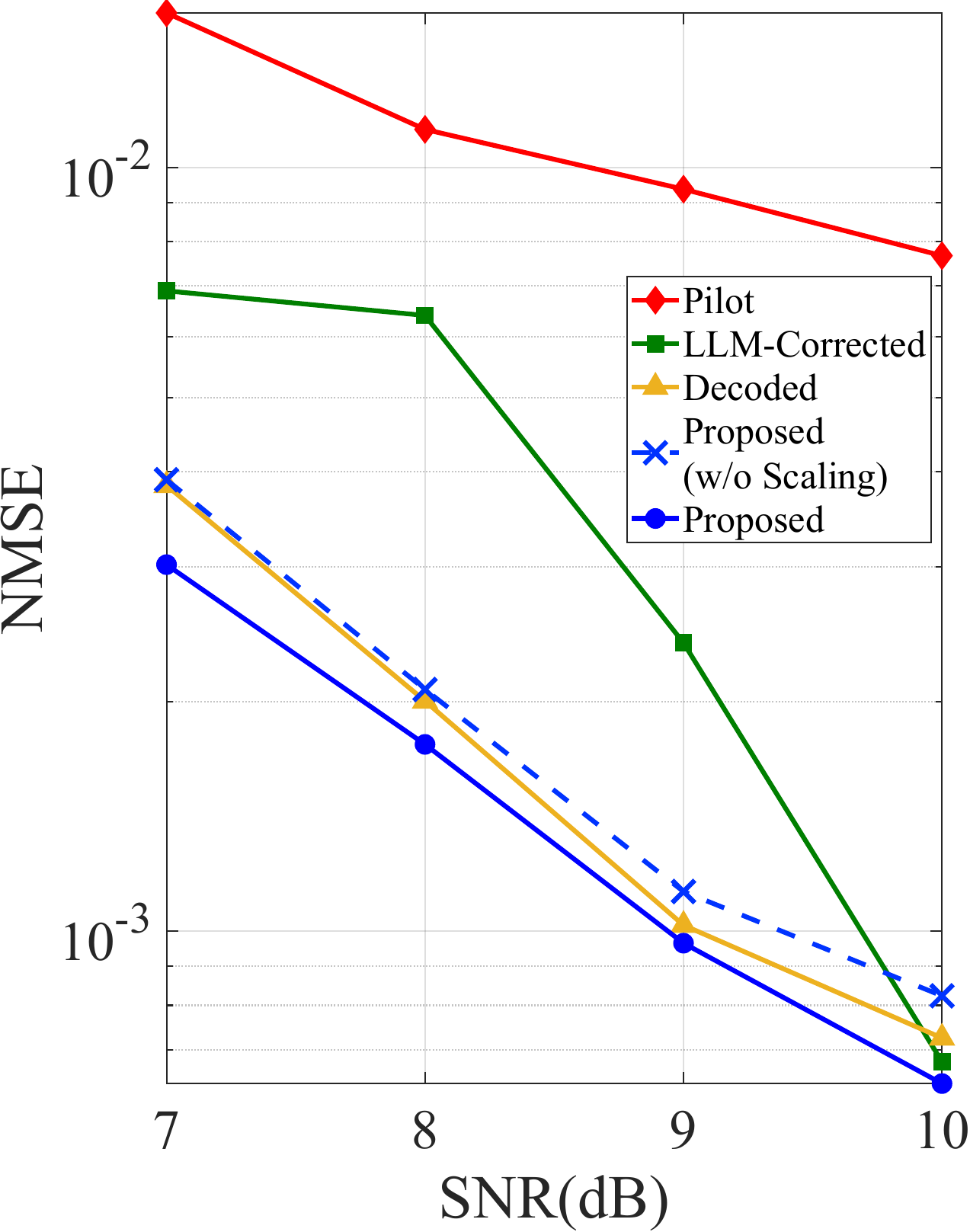}
        \caption{}
        \label{fig:nmse}
    \end{subfigure}
    \hfill
    \begin{subfigure}[t]{0.48\linewidth}
        \centering
        \includegraphics[width=\linewidth]{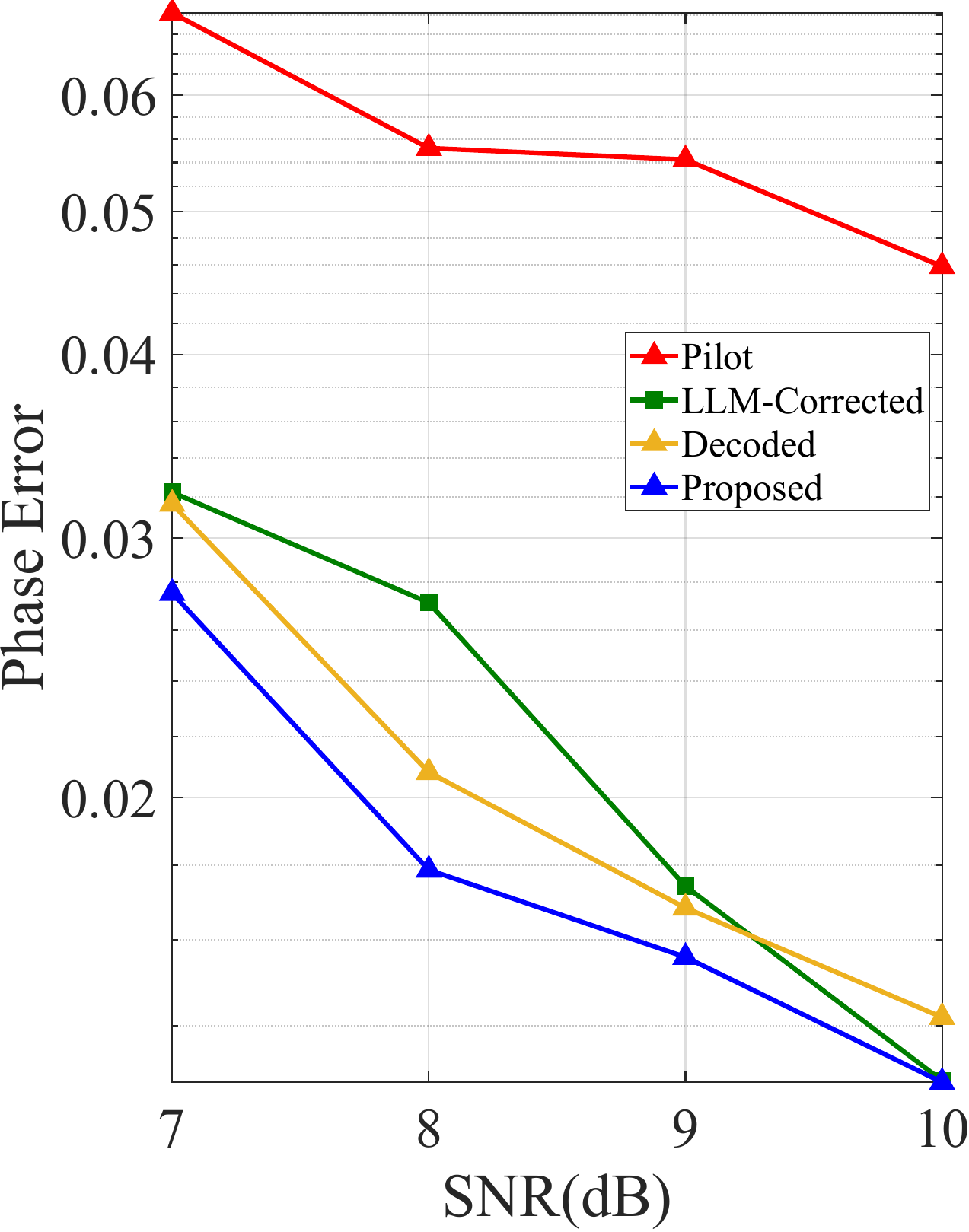}
        \caption{}
        \label{fig:phase}
    \end{subfigure}
    \caption{Performance comparison of different channel estimation schemes: (a) NMSE and (b) phase error in various SNRs.}
    \label{fig:perf_compare}
    \vspace{-3mm}
\end{figure}

\vspace{-2mm}
\subsection{Results and Discussion}
Table~\ref{tab:selection} presents the analysis of the proposed semantic pilot selection method. At a signal-to-noise ratio (SNR) of 10 dB, the method achieves a selection ratio of 96.37\%, indicating that the vast majority of the payload is utilized as pilot signals. Specifically, 97.29\% of correctly decoded symbols are selected as the semantic pilot. Furthermore, the semantic pilot symbols achieve an error-free rate of 99.94\%. Despite occasional LLM correction failures, these results demonstrate the robustness of our selection method. This effective identification of reliable symbols is critical to improving channel estimation accuracy.

The performance comparison between the proposed method and baseline channel estimation schemes is provided in Fig.\ref{fig:perf_compare} and Table\ref{tab:ber}. In particular, Fig.~\ref{fig:nmse} shows that the proposed method with magnitude scaling consistently achieves the lowest NMSE across all tested SNRs. This result demonstrates that the magnitude scaling successfully corrects the bias that remains after phase refinement, highlighting the effectiveness of the proposed two-step approach.
The phase error is particularly critical in QPSK modulation, since it directly affects the symbol detection accuracy. As illustrated in Fig.\ref{fig:phase}, the proposed method yields the lowest phase error among all baseline schemes. This reduction in phase error directly contributes to the observed improvements in BER, as summarized in Table\ref{tab:ber}. At an SNR of 9~dB, the proposed method outperforms the conventional pilot-only baseline by up to 7.85\% in BER. These results validate that leveraging semantic information is a highly effective strategy for improving channel estimation performance.


\begin{table}[t]
\caption{Performance analysis of the semantic pilot selection.}
\label{tab:selection}
\centering
{
\footnotesize
\begin{tabular*}{\columnwidth}{@{\extracolsep{\fill}} l cccc}
\toprule
\textbf{Metric} & \multicolumn{4}{c}{\textbf{SNR (dB)}} \\
\cmidrule(l){2-5}
& \textbf{7} & \textbf{8} & \textbf{9} & \textbf{10} \\
\midrule
Reliability (\%) & 98.86 & 99.08 & 99.37 & 99.94 \\
Detection Rate (\%) & 86.48 & 90.94 & 95.11 & 97.29 \\
Selection Ratio (\%) & 84.25 & 88.29 & 93.69 & 96.37 \\
\bottomrule
\end{tabular*}
}
\vspace{-2mm}
\end{table}

\begin{table}[t]
\caption{BER comparison at different SNRs. The best result in each SNR is in \textbf{bold}, and the second-best is \underline{underlined}.}
\label{tab:ber}
\centering
{ 
\footnotesize
\setlength{\tabcolsep}{3pt} 

\begin{tabular*}{\columnwidth}{@{\extracolsep{\fill}} l cccc}
\toprule
\textbf{Method} & \multicolumn{4}{c}{\textbf{SNR (dB)}} \\
\cmidrule(l){2-5}
& \textbf{7} & \textbf{8} & \textbf{9} & \textbf{10} \\
\midrule
Pilot         & 2.55e-2 & 1.63e-2 & 8.41e-3 & 5.68e-3 \\
Decoded       & \underline{2.41e-2} & \underline{1.55e-2} & \underline{7.77e-3} & 5.26e-3 \\
LLM-Corrected & 2.52e-2 & 1.69e-2 & 8.16e-3 & \underline{5.25e-3} \\
Proposed      & \textbf{2.39e-2} & \textbf{1.54e-2} & \textbf{7.75e-3} & \textbf{5.24e-3} \\
\bottomrule
\end{tabular*}
}
\vspace{-3mm}
\end{table}

\vspace{-2mm}
\section{Conclusion}
In this paper, we proposed a semantic pilot design for data-aided channel estimation. The semantic pilot is identified by comparing the initially decoded text with its LLM-corrected version. Simulation results demonstrated that the proposed method outperforms the conventional pilot-only estimation and other data-aided methods, achieving the lowest NMSE, phase error, and BER. As the proposed method operates independently of conventional data-aided techniques that rely on statistical information, it can be integrated with them to achieve complementary improvements. Future work targets multiple-input multiple-output systems and real-time deployment via lightweight LLMs.

\section{Acknowledgment}
This work was supported in part by Institute of Information \& communications Technology Planning \& Evaluation (IITP) under 6G·Cloud Research and Education Open Hub (IITP-2025-RS-2024-00428780) grant funded by the Korea government (MSIT), and in part by IITP grant funded by the Korea government (MSIT) (No.RS-2024-00404972, Development of 5G-A vRAN Research Platform), and in part by IITP-Information Technology Research Center (ITRC) grant funded by the Korea government (Ministry of Science and ICT) (IITP-2025-2021-0-02048).

\bibliographystyle{IEEEbib}
\bibliography{refs}

\end{document}